\documentclass[10pt,a4paper]{article}
\usepackage[utf8]{inputenc}
\usepackage[T1]{fontenc}
\usepackage{amsmath}
\usepackage{amsfonts}
\usepackage{amssymb}
\usepackage{graphicx}
\usepackage{siunitx}
\DeclareSIUnit\angstrom{\text {Å}}
\usepackage{xcolor}
\usepackage{authblk}
\usepackage{float}
\usepackage{placeins}

\usepackage[left=2.00cm, right=2.00cm, top=2.00cm, bottom=2.00cm]{geometry}

\title{Texture tomography with high angular resolution utilizing sparsity}

\author[a]{Mads Carlsen}
\author[a]{Florencia Malamud}
\author[b,c]{Peter Modregger}
\author[d]{Anna Wildeis}
\author[d]{Markus Hartmann}
\author[d]{Robert Brandt}
\author[a]{Andreas Menzel}
\author[a,e,f]{Marianne Liebi}

\affil[a]{Paul Scherrer Institut, 5232 Villigen PSI, Switzerland}
\affil[b]{Physics department, University of Siegen, 57072 Siegen, Germany}
\affil[c]{Center for X-ray and Nano Science CXNS, Deutsches Elektronen-Synchrotron DESY, 22607 Hamburg, Germany}
\affil[d]{Mechanical Engineering Department, University of Siegen, 57072 Siegen, Germany}
\affil[e]{Institute of Materials, Ecole Polytechnique Fédérale de Lausanne,1015 Lausanne, Switzerland}
\affil[f]{Department of Physics, Chalmers University of Technology, SE-412 96 Gothenburg, Sweden}

\date{September 23 2024}

\DeclareMathOperator{\argmin}{argmin}
\newcommand{\noleft}{\left.\kern-\nulldelimiterspace}

\begin{document}

\maketitle

\begin{abstract}
    We demonstrate a novel approach to the reconstruction of scanning probe x-ray diffraction tomography data with anisotropic poly crystalline samples. The method involves reconstructing a voxel map containing an orientation distribution function in each voxel of an extended 3D sample. This method differs from established approaches by not relying on a peak-finding step and is therefore applicable to sample systems consisting of small and highly mosaic crystalline domains that are not handled well by existing methods. This is achieved by reconstructing an orientation distribution function in each voxel rather than the orientation field. Samples of interest include bio-minerals and a range of small-grained microstructures common in engineering metals. By choosing a particular kind of basis functions, we can effectively utilize non-negativity in orientation-space for samples with sparse texture. This enables us to achieve stable solutions at high angular resolutions where the problem would otherwise be underdetermined. We demonstrate the new approach using data from a shot peened martensite sample where we are able to map the twinning micro structure in the interior of a bulk sample without resolving the individual lattice domains. We also demonstrate the approach on a piece of gastropod shell with a mosaic micro structure. The results suggest that by utilizing the sparsity of the texture, the experiment can be carried out in a simpler geometry using only a single rotations axis.
\end{abstract}

\section{Introduction}

Understanding the texture of crystalline materials is central to a wide range of phenomena in materials science. While electron backscatter diffraction (EBSD) has overtaken x-ray diffraction methods to become the main tool for texture characterization over the last twenty years, there are scientific questions where the high penetrating power of x-rays is needed to find the answers. As EBSD is a surface imaging technique, its analysis is susceptible to be biased by 2D effects as it cannot see the local environment hidden in the third dimension\cite{JUULJENSEN2020,Knipschildt_2023}. While full 3D maps can be obtained by the serial sectioning technique, such imaging is inherently destructive and precludes in-situ characterization. Therefore, x-ray diffraction techniques play an irreplaceable role in the characterization of micro structure whenever bulk properties are of interest. While traditional x-ray diffraction techniques probe the volume-averaged structure of extended samples, a range of tomographic techniques have been developed that yield spatially resolved information.

\begin{figure}
    \centering
    \includegraphics[width = 1.0\textwidth]{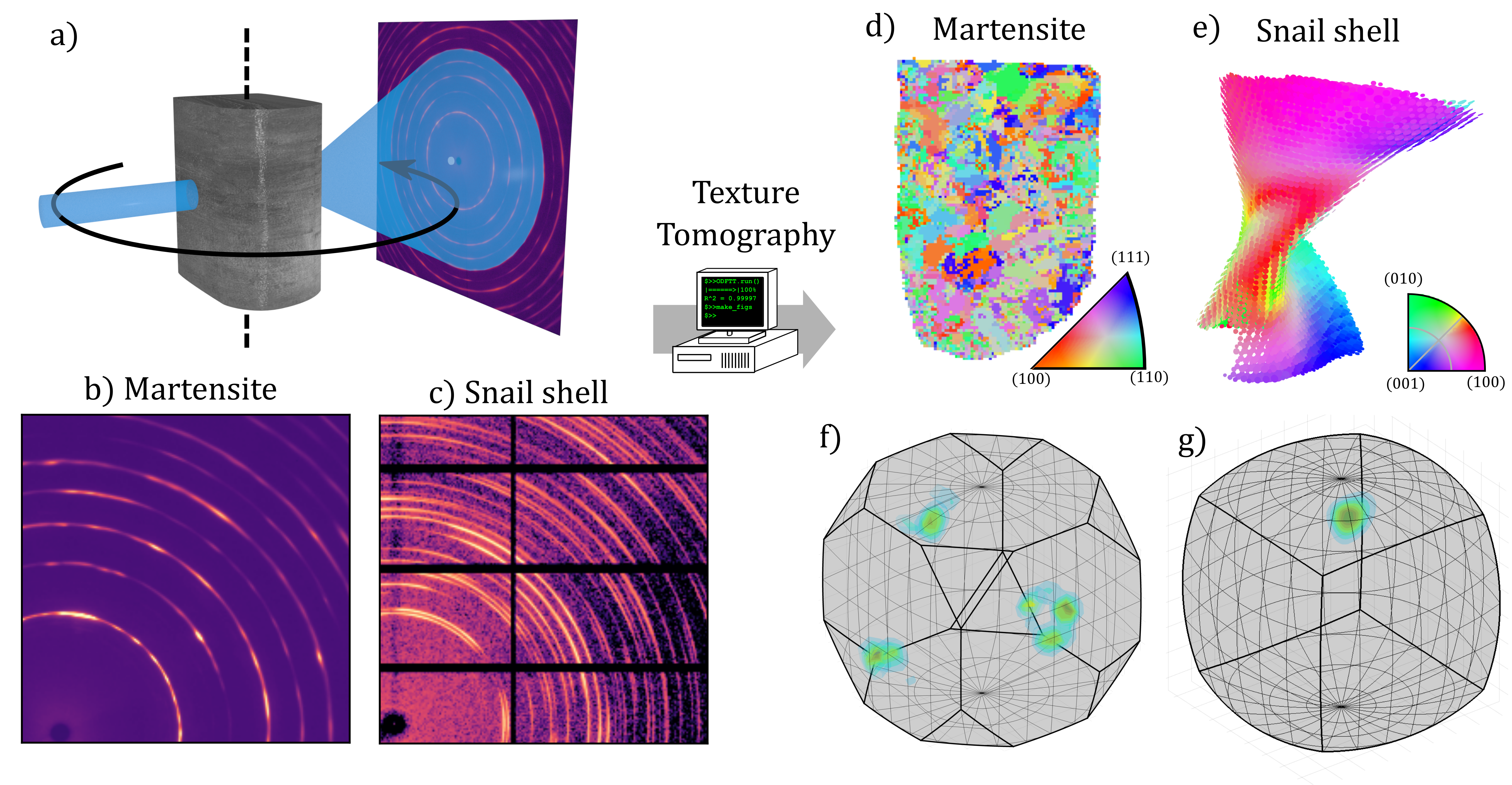}
    \caption{a) Sketch of the experimental geometry and b,c) examples of x-ray diffraction patterns from each of the two investigated samples. d,e) show the reconstructed samples as inverse pole figure maps of the main orientation and f,g) show the full orientation distribution function of a single voxel plotted as a three-dimensional density in Rodriguez-vector space in the asymmetric zone of the f) cubic and g) orthohombic crystal system of ferrite and aragonite respectively.}  
    \label{fig:figure_1}
\end{figure}

In this paper, we present a new approach for analysing the data from x-ray diffraction computed tomography (XRD-CT) experiments\cite{harding_xrd_1987, stock08, bleuet08}. XRD-CT is a synchrotron x-ray technique that works by scanning a sample through a point-focused x-ray beam and rotating the sample to record a series of x-ray diffraction sinograms. The geometry of the experiment is sketched in Fig. \ref{fig:figure_1}a). While traditional XRD-CT assumes an isotropically scattering sample, the technique proposed in this paper explicitly reconstructs the anisotropy of the sample resulting in a spatially resolved 3D map of the crystallographic texture of a polycrystalline sample.

The proposed method extends the principles of tensor tomography\cite{malecki14,liebi15,schaff15} and is formally identical to the method developed by Frewein et al.\cite{frewein_2024} called texture tomography except with a different choice of basis functions. Due to this choice of basis functions, our proposed method can effectively enforce the prior knowledge of sparsity on the solution which is necessary to ensure unique solutions at high angular resolution. As texture tomography and the method proposed in this paper reconstruct the orientation distribution function (ODF) in each voxel whereas the established tensor tomography techniques reconstruct pole figures, we introduce the abbreviations ODF-TT and PF-TT to distinguish between them.

Our method is similar to scanning three-dimensional x-ray diffraction (s3D-XRD)\cite{hayashi15,henningsson2020,bucsek_2023} which uses the same measurement approach but aims to reconstruct the fully resolved orientation field rather than the spatially averaged ODF. More generally, the proposed method falls within a large family of x-ray techniques which are collectively referred to as three-dimensional x-ray diffraction (3D-XRD)\cite{poulsen04} which contains s3D-XRD\cite{hayashi15} as well as a number of full-field techniques\cite{Poulsen_2001,suter_2006,Schmidt_2008,Ludwig_2008,vigano_2016,oddershede_2019}. However, the tensor/texture tomography approaches fundamentally differ from the established techniques that rely on spatially resolving the smallest coherent lattice domains of the sample (be it grains, sub-grains, or twin domains) by indexing distinct diffraction spots that can be assigned to individual grains. Rather than this, the tensor-tomography methods work with smeared out diffraction features that Debye-Scherrer rings with a continuous intensity variation along the detector azimuth (as shown in Fig. \ref{fig:figure_1}b,c) and reconstructs the spatially averaged texture over many crystalline domains. Therefore, the techniques are suitable for a range of small-grained and highly deformed materials such as bone-apatite\cite{gruenewald20, gruenewald23} and cold-worked metals\cite{carlsen_2024}.

ODF-TT overcomes some of the weaknesses of other x-ray scattering-tomography techniques and opens up possibilities for investigating  samples, not covered by existing techniques. By reconstructing the full ODF using a grid based expansion of the ODFs, non-negativity in orientation-space can be enforced, overcoming certain ambiguities of the inversion problem. Furthermore, by enforcing the lattice symmetries we overcome the `missing wedge' problem of PF-TT\cite{schaff15, nielsen_2024_a} allowing experiments to be carried out in a simpler geometry without a second rotation stage which is commonly used in PF-TT. We present the mathematical approach to the reconstruction and apply the method to a 2D XRD-CT dataset from a piece of shot-peened martensite and to a full 3D tensor-tomography dataset of a piece of gastropod shell.

\section{Results}

To test the feasibility of ODF-based TT, we show results using two separate experimental datasets. One is a piece of martensitic steel which is a sample system of high scientific interest and with a well known twinned microstructure that has so far not been possible to map with 3D-XRD approaches. The other is a biomineral sample where a full 3D tensor tomography data set is available to allow us to compare the performance of ODF-TT with PF-TT. Fig. \ref{fig:figure_1}b,c) show examples of raw detector frames from each of these two data sets. Both data sets display strongly anisotropic Debye-Scherrer rings but lack easily identifiable separate diffraction spots.

\begin{figure}
    \centering
    \includegraphics[width = 0.5\textwidth]{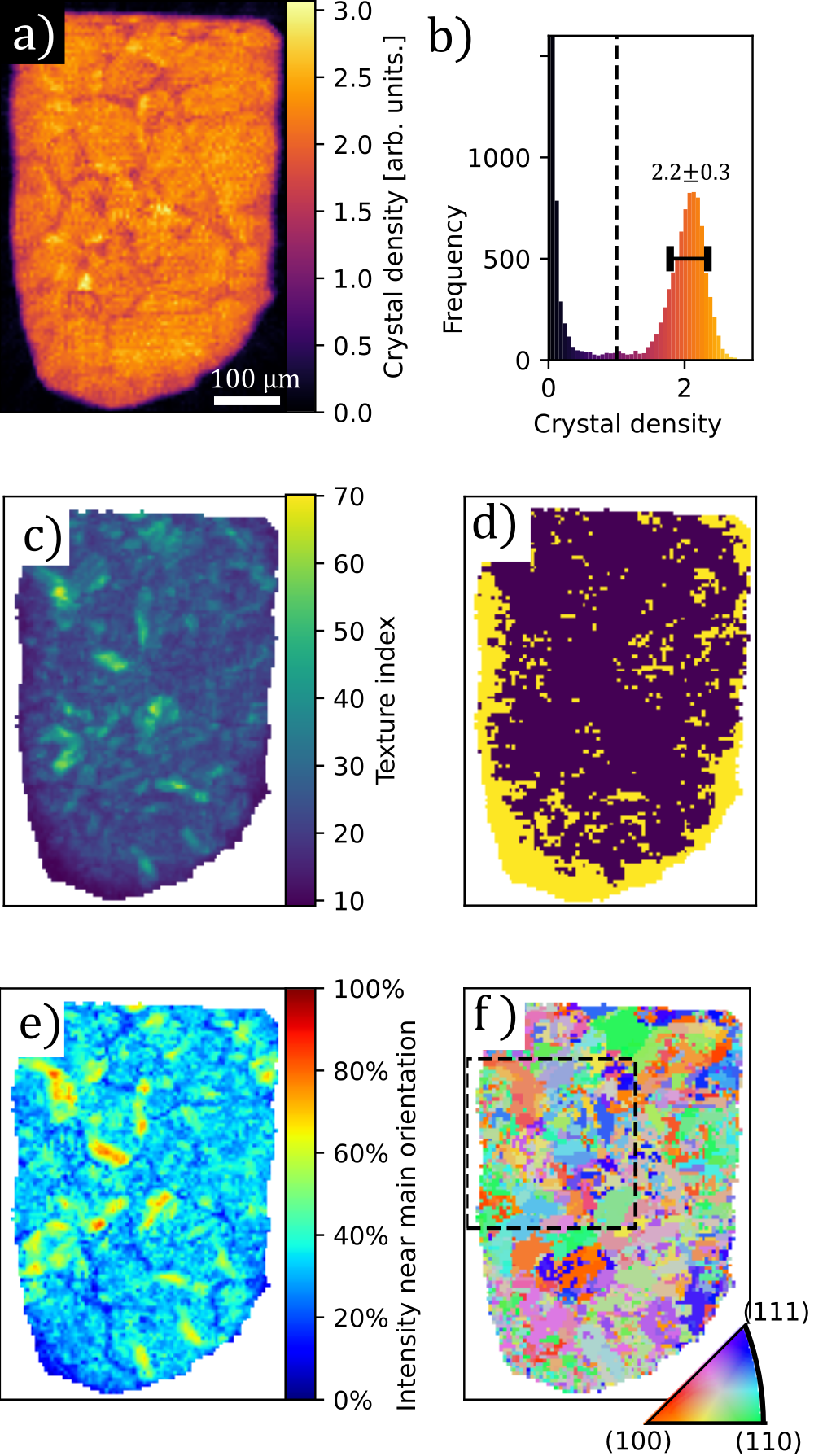}
    \caption{Reconstructed tomogram of the steel-sample. a) crystal density, b) histogram of the reconstructed density values. The dashed line marks the threshold used to generate the mask distinguishing sample from the air. c) approximate texture index computed from the standard deviation of the reconstructed coefficients of each pixel d) a binarized map showing regions with texture-index lower than 20 in yellow e) fraction of the ODF-density that falls within a sphere of radius 20$\si{\degree}$ of the main-orientation. f) Inverse pole figure map of the main orientation determined as the grid orientation corresponding to the largest coefficient in each voxel. }
    \label{fig:steel_1}
\end{figure}

Figure \ref{fig:steel_1} shows a number of quantities computed from the reconstruction of the martensite data set. In Fig. \ref{fig:steel_1}a,b) we see that the shape of the sample is well reconstructed and that the reconstructed density is fairly homogeneous as expected for the sample which consists almost entirely of a single crystalline phase (ferrite). In Fig. \ref{fig:steel_1}c) we display the texture index of the individual voxels. The texture index is a measure of the sharpness of the texture which is equal to one for a random (uniform) texture and increases towards infinity for a single crystal. We observe that the shot-peened surfaces (that is, all surfaces except the straight edge at the top of the figure) display a markedly lower texture index than the rest of the sample. This observation is highlighted in Fig. \ref{fig:steel_1}d) by thresholding regions of low texture index.

We determine a main orientation in each voxel (see Methods section \ref{sec:analysis} for details of the computation). This main orientation is plotted as an inverse pole figure map in Fig. \ref{fig:steel_1}f). The microstructure of martensite consists of narrow laths of ferrite with sub-micron thickness; smaller than the resolution of the experiment. The lattice orientations in the individual laths are related to their neighbours through transformation twinning from the austenite parent phase and therefore follow orientation relationships which causes the same orientations to be repeated in nearby grains. The domains observed in the reconstruction are thus not the individual laths but rather the outlines of the former austenite parent grains before transformation into martensite. This is demonstrated by Fig. \ref{fig:steel_1}e) which shows that only about half of the total intensity can be attributed to a main orientation. Fig. \ref{fig:steel_1}e) shows the main orientation plotted as an inverse pole figure map. Some large domains of sizes up to 50\si{\micro m} are present, but regions with domains down to the resolution of the experiment are also observed.

\begin{figure}
    \centering
    \includegraphics[width = 0.5\textwidth]{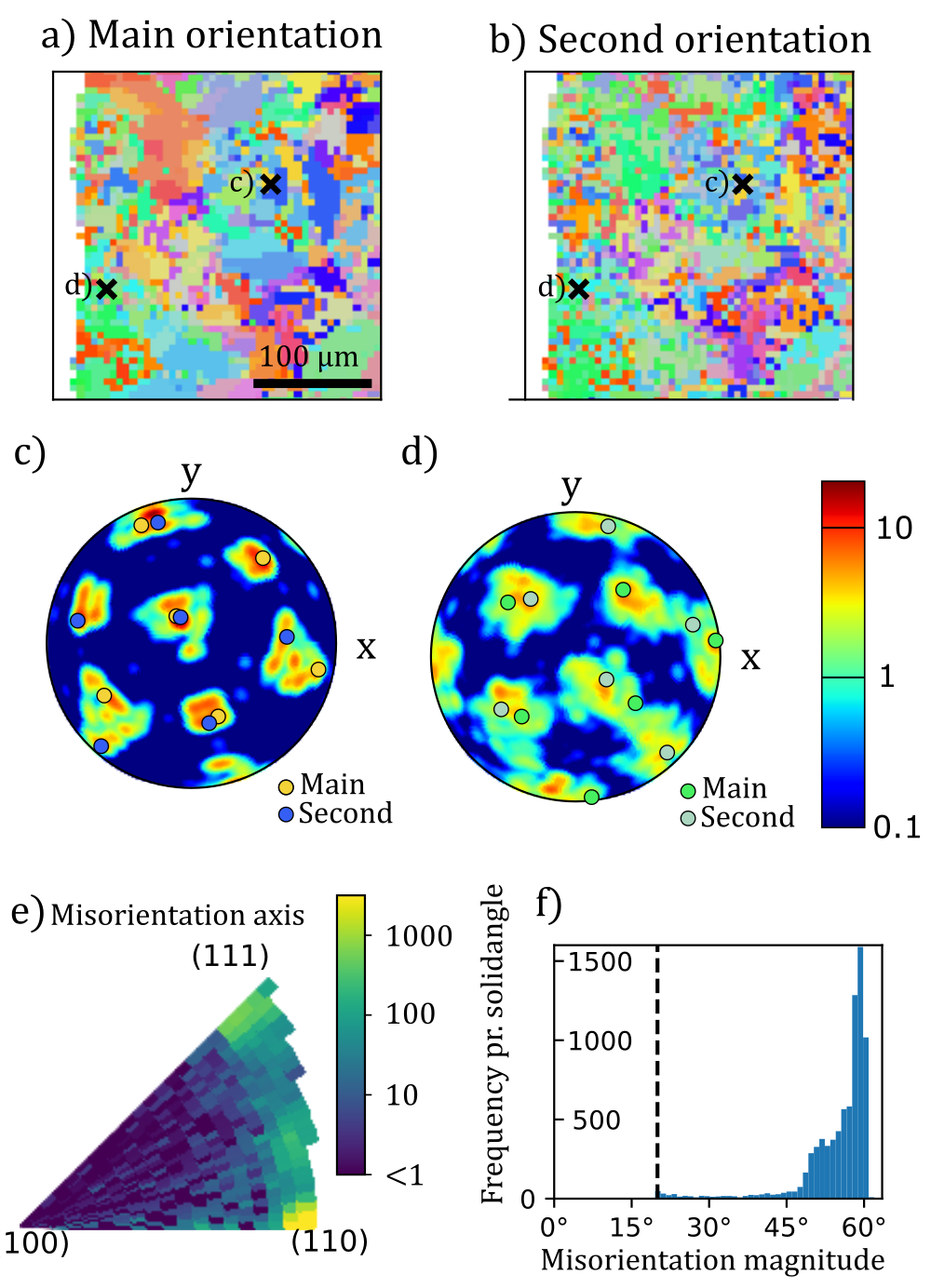}
    \caption{Zoomed-in view of the region of the martensite sample marked in Fig. \ref{fig:steel_1}f). Inverse pole figure maps of the a) primary and b) secondary orientations. c,d) show the \{110\} pole figures of two 4 by 4 pixel regions, marked in a) and b). Overlaid on these pole figures are the poles corresponding to the main orientations colored by the corresponding IVP color. Histograms of the e) direction of the misorientation axis between the primary and secondary orientations in lattice coordinates and f) the magnitude of the misorientation.
    }
    \label{fig:steel_2}
\end{figure}

In Fig. \ref{fig:steel_2} we zoom in and provide a view of the local texture. After subtracting the components within a 20\si{\degree} radius of the main orientation, we can compute a second orientation using the same approach. The local texture is in most pixels dominated by three distinct orientations known as Bain groups that appear at large misorientation angles to each other which is typical of martensite. For the two selected regions shown in figure \ref{fig:steel_2}c,d), we see that the region in the interior shows typical martensite texture while the region near the peened surface has more diffuse intensity away form the major peaks. This is what causes the lower texture index and is interpreted to be due to deformation twinning, stemming from the peening process.

To quantify the twinning in the sample, we calculate the misorientation between the primary and secondary orientations for each voxel. Fig.\ref{fig:steel_2} e) shows the distribution of the direction of the misorientation axis in the lattice coordinates. We see that the misorientation is mainly around $\langle 110 \rangle$ directions and directions close to, but not quite aligned with $\langle 111 \rangle$. Fig.\ref{fig:steel_2} f) shows the magnitude of the misorientation. The way we calculated the secondary orientations precludes the observation of misorientations with magnitude smaller than 20\si{\degree}, but beyond this, almost no misorientations between 20\si{\degree} and 45\si{\degree} are detected. The found directions and magnitudes of the misorientations are consistent with the three Bain groups in martensitic transformation twinning and follows the expected orientation relationships\cite{Kurdjumow_1930Z, guo_2004}.

\begin{figure}
    \centering
    \includegraphics[width = 0.5\textwidth]{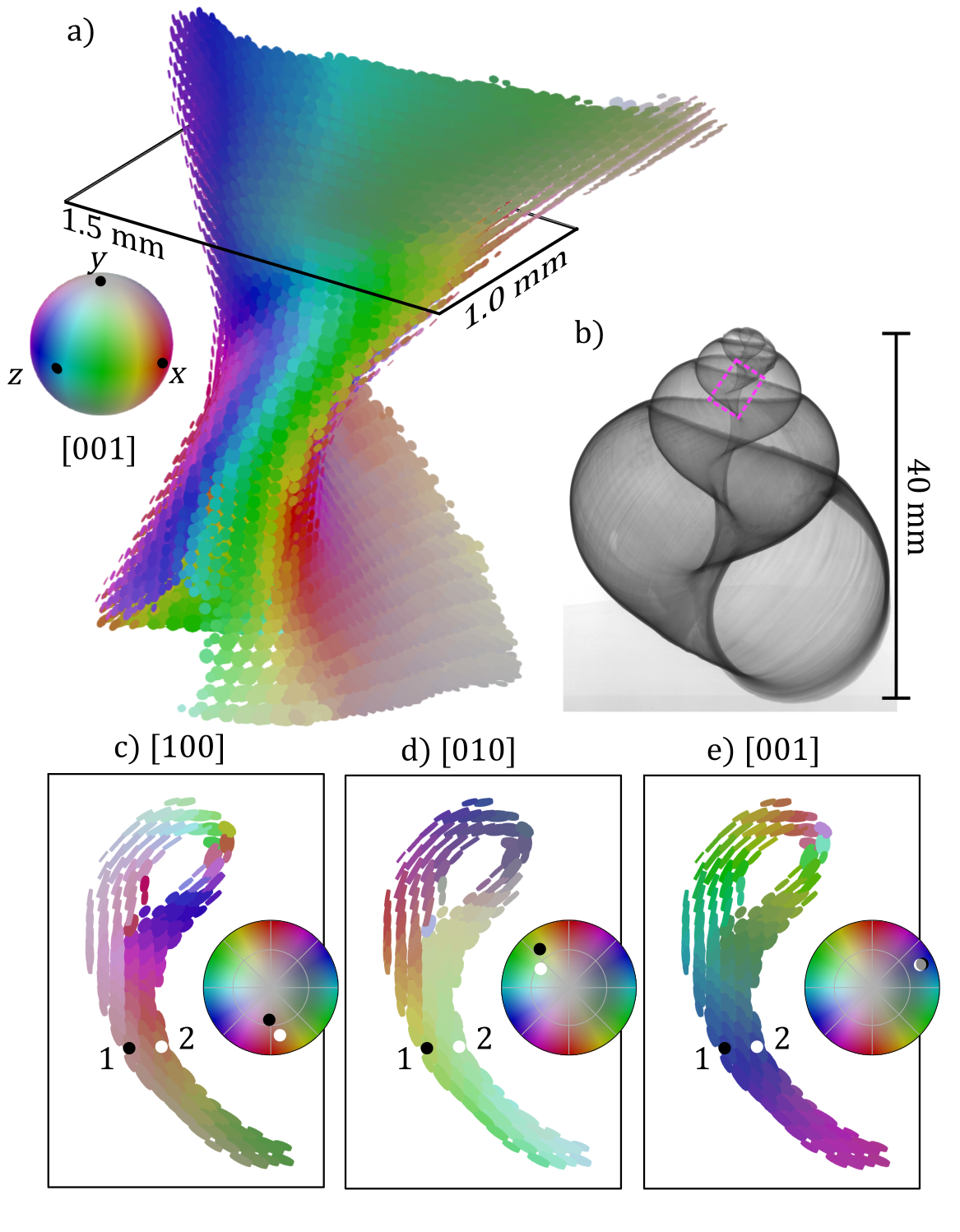}
    \label{fig:snail_1}
    \caption{3D renders of the reconstructed snail-shell sample using ODF-TT. a) Shows the full reconstructed volume rendered as flat cylinders with the face aligned with the c-axis of the main orientation and where the size of the cylinders are proportional to the reconstructed density. The cylinders are coloured according to the direction of the c-axis. b) Radiograph of a Roman snail shell showing the approximate location from where the sample was extracted. c-d) single slice of the reconstructed volume at the location marked with a black rectangle in a) coloured according to the direction of the a, b and c axis respectively. For two select voxels on each side of the columnar wall, labeled 1 and 2, the direction of each of the three primary axes is plotted showing an abrupt change of about 31$\si{\degree}$ in the direction of the a and b axes. }
\end{figure}

The other sample investigated is a piece of gastropod shell and consists predominately of aragonite, a calcium carbonate polymorph belonging to the orthogonal crystal family. The local micro structure is dominated by a single orientation with a mosaic spread of around 20\si{\degree} FWHM (See Fig. \ref{fig:figure_1}g)). Fig. \ref{fig:snail_1} a) shows the direction of the orthorhombic $c$-axis in the mean orientation. The $c$ axis is aligned with the surface normal of the columellar wall, which is typical for aragonite shells in mollusks\cite{boggild_1930}. Fig. \ref{fig:snail_1} c,d,e) show a single slice of the reconstruction displaying the direction of the three orthorhombic axes. From these figures we can see that where the columnar wall folds back on itself after wrapping around the umbilical, it consists of two distinct layers with different crystallographic orientations that are related by a rotation about the common $c$-axis. These two layers are part of two distinct whorls of the shell and were as such grown at different points in the snails lifetime.

The gastropod shell data set can also be reconstructed using established techniques from PF-TT. To give a fair comparison, we use a set of basis functions similar to the ones used in the ODF-TT reconstructions consisting of symmetric Gaussian functions placed on a grid of directions covering the half-sphere\cite{nielsen_2024_a} which can also benefit from the non-negativity constraint and sparse texture. To facilitate comparison between the two reconstruction methods, we chose to reconstruct the 200 Bragg peak as it is the only peak in the dataset which is to parallel one of the main crystallographic axes and has multiplicity 2, which makes it possible to define a main direction of the scattering. The reconstructions were performed using the software package \texttt{mumott}\cite{nielsen2023b}.

\begin{figure}
    \centering
    \includegraphics[width = 0.5\textwidth]{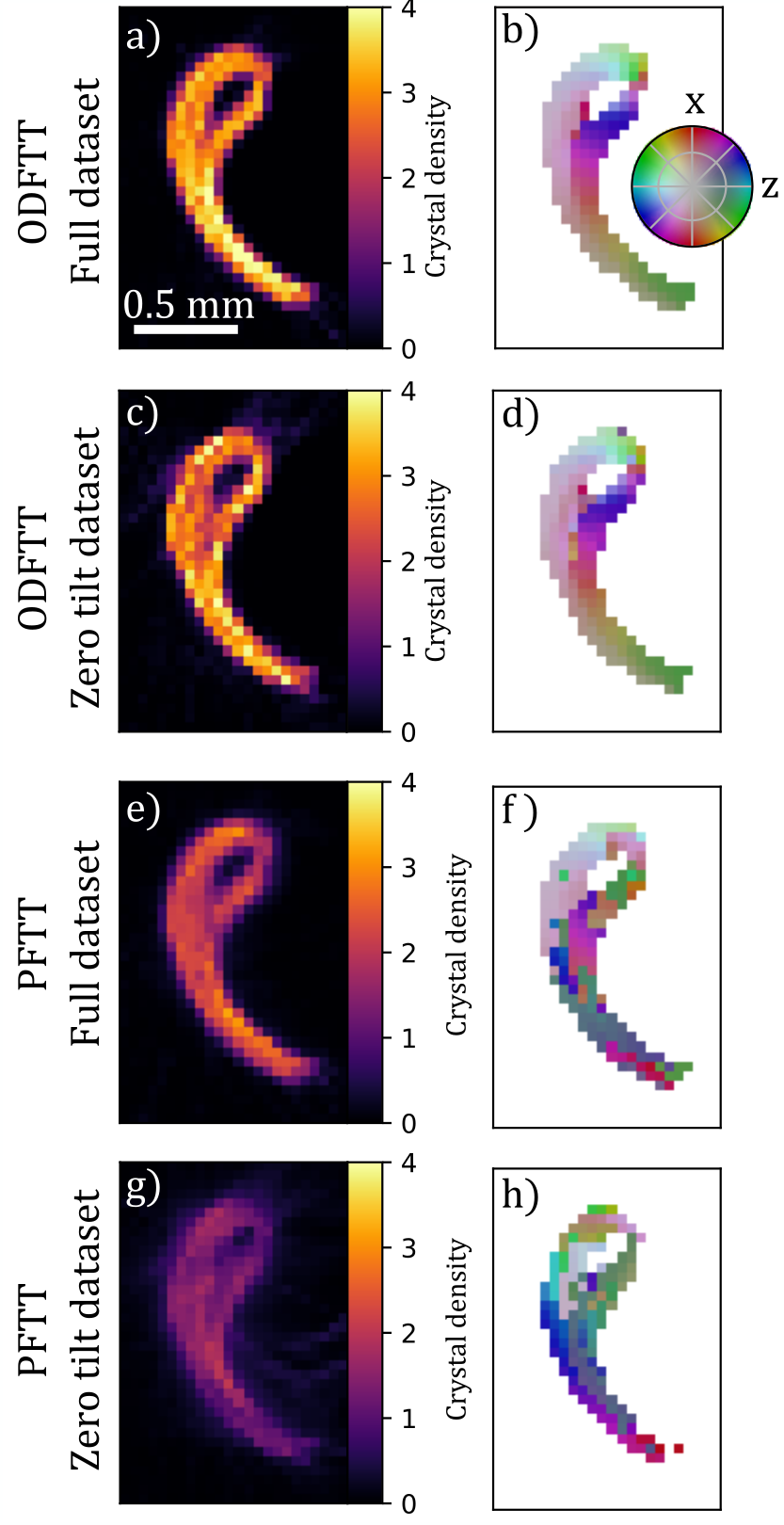}
    \caption{Comparison of reconstructions using ODF-TT and PF-TT of the gastropod shell sample. a,c,e,g) show a single slice, orthogonal to the shell axis, of the reconstructed crystal density while b,d,f,h) shows the main orientation of the (100) crystallographic direction using a color coding. a,b) are from ODF-TT reconstructions using the full dataset with tilts up to 45\si{\degree} and c,d) from ODF-TT using only the zero-tilt part. e,f) are from a PF-TT reconstruction using the full data set g,h) are from a PF-TT reconstruction made using only the zero-tilt part of the data set.}
    \label{fig:snail_2}
\end{figure}

Fig. \ref{fig:snail_2} shows a comparison between the ODF-TT reconstructions and the PF-TT reconstructions. OFT-TT reconstructs a smoothly varying direction both with the full and zero-tilt data sets except where the columnar wall touches itself after folding around the umbilical. The direction found from the PF-TT reconstructions on the other hand has sudden variations along the columnar wall and appears more erratic. With the full dataset, the PF-TT direction is close to the ODF-TT directions at many points in the sample, especially near the center where the sample is thicker, but differs significantly at the narrower features. This is consistent with the type of `missing wedge` artifacts common in PF-TT where flat features are difficult to reconstruct when the projections orthogonal to the normal of the plane have not been measured. With only the zero tilt data, there are no longer any clear correlation between the direction of the PF-TT reconstruction and of the ODF-TT reconstructions. The supporting information \ref{fig:supplement_3} shows a comparison of individual pole figures, and we see that while the main peak is also reconstructed by PF-TT, there are additional peaks of similar amplitude that obfuscate the main orientation.

\section{Discussion}

Different versions of volumetric grain-mapping synchrotron x-ray techniques have existed for more than twenty years\cite{Poulsen_2001}. However, they are mostly limited to either materials with low intragranular misorientation and strain or samples with a small number of grains. Recent advances have targeted larger strains by utilizing a focused beam\cite{hayashi15}, conical slits\cite{hayashi_2023}, or by improved computational approaches\cite{henningsson_2024}. Still these approaches have so far only been demonstrated on fairly simple grain structures. Martensite and other highly strained phases, which Hayashi et al. refers to as `invisible phases'\cite{hayashi_2023_b}, have thus far not been possible to reconstruct with 3D-XRD techniques. 

Vigano et al.\cite{vigano_2016, Vigano2024} showed that the 3DXRD problem could be linearized by writing the problem as a search for a distribution in six-dimensional orientation-position space instead of an orientation field. However, this approach has not been applied to the scanning beam geometry and still depends on an initial peak-finding step. Dark-field x-ray microscopy\cite{simons_2015} and differential aperture x-ray microcopy\cite{yang2004, Larson_2013} have both been used to image highly deformed microstructures in the bulk of extended samples\cite{Yildirim_2023,yu_2023} but work with smaller volumes and at higher angular and spatial resolution.

The aim of ODF-TT is less ambitious than many of the existing 3D-XRD techniques, as it does not aim to reconstruct the individual grains or the strain field and has lower angular resolution. However, compliments these techniques due to its simplicity and ability to study less well ordered materials such as twinned and highly strained microstructures that are of clear technological importance.

For both samples investigated in this study, the reconstruction problem is underdetermined with around 240 million and 1.8 billion degrees of freedom for the martensite and snail-shell samples respectively compared to just 15 million and 600 million data points. Due to the sparsity of the texture however, a large majority of the expansion coefficients are zero in the converged solutions with only 0.6 million and 6 million non-zero coefficients respectively. The sparsity of the sample texture is thus critical to ensure convergence to a stable solution and without the non-negativity constraint the method suffers from high-frequency over fitting artifacts. (See supplementary information Fig. \ref{fig:supplement_1}) We note however that the sample-averaged texture does not need to be sparse, only the voxel-averaged texture over a length scale given by the resolution of the experiment.

The output of an ODF-TT reconstruction is an ODFs for each voxel in the reconstructed volume, and both analysis and visualization of the reconstruction is challenging. For a sample such as the gastropod snail, where a main orientation can be defined, the information contained in ODFs can be reduced to a small number of derived quantities, such as the orientation field, density and mosaic spread. For more complex and not fully resolved microstructures such as the martensite, more complicated analysis is required to interpret of the reconstructions. This involves processes such as martensite packet and parent grain determination\cite{Niessen_2022}. While the field of texture analysis provides methods for these tasks, they are typically designed for different settings. In electron back scatter diffraction(EBSD) images, the orientation field is well resolved and the task involves grouping neighboring pixels based on their relative misorientations. Traditional x-ray texture mapping on the other hand relies on a high-angular resolution texture map of the entire sample to be known. In summary, existing algorithms need to be adapted for in-depth analysis of ODF-TT reconstructions. 

\section{Conclusion}

We have demonstrated a novel approach for tomographic 3D mapping of crystallographic texture in extended samples using a method similar to pole-figure based tensor tomography. By utilizing grid based basis functions and leveraging sparsity, we can overcome the inherent ambiguities of the inversion problem. Our proof-of-principle experiments show that a reconstruction can be achieved even for highly under-constrained problems by applying a non-negativity constraint in samples with locally sparse textures.

Unlike for PF-TT, our method achieves good reconstructions with measurements using only a single rotation axis. This means that both experiments and reconstructions can be carried out in a slice-by-slice manner which enables fast single-slice experiments with measurement times on the order of minutes rather than hours. This simplifies the experimental procedure by not requiring a second rotations stage, fascilitating in-situ experiments using various existing sample environments at a range of synchroton end stations.

Our findings suggest that ODF-TT with grid-type basis functions will be able to extend the range of samples that can be characterized with existing 3D-XRD and PF-TT techniques. This includes twinned and deformed metal microstructures and broadly mosaic biominerals.

\section*{Acknowledgments}

 We acknowledge the Paul Scherrer Institute, Villigen, Switzerland for provision of synchrotron beamtime at the cSAXS beamline of the SLS and we would like to thank C. Appel and M. Olson for assistance during the beamtime. We acknowledge DESY (Hamburg, Germany), a member of the Helmholtz Association HGF, for the provision of experimental facilities. We thank Dr. M. Shakoorioskooie and the NEUTRA beamline at SINQ for access to their x-ray radiography equipent. Parts of this research were carried out at the beamline P21.2 of PETRA III.
 Funding information: MC has received funding from the European Union’s Horizon 2020 research and innovation program under the Marie Skłodowska-Curie grant agreement No 884104. ML has received funding from  the European research council (ERC-2020-StG 949301 MUMOTT).

\section{Author contributions}
M.C.: Conceptualization, Writing -- original draft, Software, Formal analysis, Investigation, Visualization, Methodology
F.M.: Formal analysis, Visualization, Writing -- review \& editing
P.M.: Investigation, Resources, Formal analysis, Visualization, Writing -- review \& editing
A.W.: Resources
M.H.: Resources
R.B.: Resources, Supervision
A.M.: Supervision, Funding acquisition, Writing -- review \& editing
M.L.: Supervision, Visualization, Funding acquisition, Methodology, Writing -- review \& editing


\section{Method}

 For a thorough introduction to the methodology of tensor- and texture tomography, we refer to papers on the subject such as \cite{nielsen2023} and \cite{frewein_2024} that both utilize a model and geometry similar to the one applied here. The methods arise naturally by combining established approaches from texture inversion and computed tomography. We first present some of the key results of texture inversion and then describe how the tomographic problem is constructed and solved.

\subsection{Texture inversion} 
While tensor tomography was originally applied for small angle x-ray scattering\cite{liebi15,schaff15}(SAS-TT), its extension to wide-angle Bragg scattering is straightforward\cite{gruenewald20, carlsen_2024}. In SAS-TT, the reconstructed quantity is called the reciprocal space map. For crystalline systems displaying scattering only in discrete shells corresponding to the length of reciprocal lattice vectors of the lattice, the reciprocal space map on such a sphere is called a pole figure. These pole figures have to follow symmetries corresponding to the rotational symmetries of the crystal lattice. However, the symmetries are not fully realized in the individual pole figures but result in correlations between different pole figures. This motivates the reconstruction of a different quantity, the ODF which describes the anisotropy of all Bragg peaks simultaneously and allows the rotational symmetries to be included. 

The reconstruction of the ODF from x-ray diffraction measurements is called texture inversion. Like computed tomography, texture inversion is an inverse problem that revolves around reconstructing a $3$-dimensional function from a series of measurements of $2$-dimensional projections. In computed tomography, this projection is defined by the well-known Radon- or x-ray transform in Cartesian coordinates. In texture inversion, the projection operation is given by a similar integral equation that relates a pole figure defined on the two dimensional space of directions to an ODF defined on the three dimensional space of proper rotations. 

Texture inversion problems are most commonly analyzed using the harmonic method, in which the ODF is expanded in a series of generalized spherical harmonics\cite{roe_1965, Bunge_book_1}. One of the most important properties of the harmonic expansion in this regard is that the projection operator applied to the generalized spherical harmonics of a given order $\ell$ are projected onto spherical harmonics of the same order in the spherical harmonics expansion of pole figures. This property allows splitting the inversion into independent problems for each order $\ell$ and provides explicit quadrature equations to determine the coefficients of the ODF when complete measurements of the pole figures are available\cite{Bunge_book_1}.

However, the harmonic method of texture inversion suffers from reconstruction artifacts termed the ghost problem\cite{matthies_1979}. This issue arises because the ODF generally consists of both even-$\ell$ and odd-$\ell$ terms in the harmonic expansion, but odd terms of the pole figures cannot be measured by x-ray diffraction due to Friedel's law which causes inversion symmetry of the measured pole figures.

To overcome this shortcoming, a different family of methods were developed starting from the late seventies. These methods expand the ODF in a series of localized functions centered on a grid of orientations that maps the asymmetric zone of the crystal lattice. These functions are typically either: (1) the indicator-functions of some partition of the asymmetric zone, resulting in a piecewise-constant approximation of the ODF\cite{ruer_1977, schaeben_1994}, (2) finite elements, which gives a piecewise linear approximation\cite{barton_2002}, or (3) spherically symmetric standard-functions\cite{schaeben_1996} such as a spherical Gaussian function used in this work.  While this change of basis does not eliminate the ghost problem, it makes it easier to enforce certain kinds of prior knowledge and to compute regularized solutions\cite{schaeben_1988} that can alleviate the problem. For sparse textures, where the ODF is close to or equal zero in extended regions of orientation space, the non-negativity constraint has been observed to resolve the inherent ambiguity of the texture inversion problem\cite{matthies_1982,dahms_1988}. 

\subsection{The tomographic inversion problem}\label{sec:model}

To write up the full forward model, we expand the pole figure of the voxel at position $x$, $y$, $z$ as a series expansion of ODF basis functions:

\begin{equation}
    O_{xyz}(g) = \sum_\mu c_{xyz\mu} O_\mu(g)
\end{equation}

where $c_{xyz\mu}$ are the unknown expansion coefficients, $g$ is a proper rotation, and $O_\mu(g)$ are the basis functions. The basis functions used here are sets of Gaussian radial functions obeying the lattice symmetry rotated to a set of grid points $g_\mu$ chosen to uniformly fill out the asymmetric zone of the respective lattice groups. The ODF defined by this expression is not properly normalized and the reconstructed quantity is actually the density of the crystalline phase times the ODF.

For each ODF basis function, we can pre-compute the corresponding pole-figure values at the points given by the geometry of the experiment. These values are stored in a set of matrices with elements:

\begin{equation}
    B^\mu_{ch}(R_i) = \mathcal{P}_{h}\{O_\mu\}(R_i^{\top}\mathbf{q}_{hc}),    
\end{equation}

where $\mathcal{P}_{h}\{\cdot\}$ is the pole figure projection operator in ODF space, $R_i$ denotes the setting of the rotation stage, and $\mathbf{q}_{hc}$ is the scattering vector measured by detector segment at the azimuthal channels labeled by $c$, and the Bragg peak labeled by $h$. These matrix coefficients can be computed with standard approaches from texture inversion and open-source software packages\cite{hielscher_2008}.

The full tomographic problem can now be written as

\begin{equation}
    I_{chjk}(R_i) = \sum_{xyz} P^{xyz}_{jk}(R_i) \sum_\mu B^\mu_{ch}(R_i) c_{xyz\mu},
\label{eq:linear_model}
\end{equation}

where $I_{chjk}(R_i)$ is the 5-dimensional dataset containing the integrated intensities and $P^{xyz}_{jk}(R_i)$ are the matrix elements of the discrete x-ray transform that transforms the voxel-grid ($x$, $y$, $z$) to the coordinates of the raster-scan grid ($j$, $k$). In a 2D geometry, the indices $y$ and $k$ can be omitted and $P$ is the Radon transform. To write the problem in this form, we have assumed that the scaling-factors that describe the relative intensity of different Bragg peaks are known beforehand and have been normalized out of the measured intensities which is needed to write the forward model in this linear form\cite{hielscher_2008}.

We rewrite the model with linear algebra syntax as

\begin{equation}
    \mathbf{I} = \mathrm{A} \mathbf{c}
\label{eq:linear_model2}
\end{equation}

by flattening the data set into a single vector $\mathbf{I}$, the expansion coefficients into a vector $\mathbf{c}$ and the products of $P^{xyz}_{jk}(R_i)$ and $B^\mu_{ch}(R_i)$ into a single matrix $\mathrm{A}$. This forms a large system of linear equations that will typically be underdetermined when the ODF-expansion is performed to high angular resolution. Here, we will focus on textured materials, where the ODF is close to zero in large regions of orientation space. This adds the prior knowledge needed to overcome the inherent ambiguities of the inversion problem. To find the solution we minimize the square residual with L1 regularization.



\begin{equation}
\begin{split}
    \mathbf{c}_{\mathrm{Opt}} &= \underset{\mathbf{c}}{\argmin} ||\mathbf{I} - \mathrm{A}\mathbf{c}||^2_2 + \lambda||\mathbf{c}||_1 \\
    &\mathrm{s.t. }\quad c_{xyz\mu} \geq 0 \quad \forall x,y,z,\mu,
\end{split}
\label{eq:optimization}
\end{equation}
where $\lambda$ is a regularization parameter and $||\cdot||_n$ denotes the Euclidean- and 1-norm for $n$ equals 2 and 1 respectively. To solve this optimization problem, we use projected gradient descent with Nesterov momentum. For both samples investigated in this paper, ODF-TT achieves good reconstructions without regularization, but L1-norm regularization is used to reduce streaking artifacts appearing in the gastropod shell sample (Supplementary material \ref{fig:supplement_3}).

Similar to texture inversion, reconstruction of tensor tomography also suffers from ambiguities that are due to the experimental difficulty of properly sampling the full range projection directions. While standard computed tomography problems can be inverted using projections measured around a single rotation axis, PF-TT requires sampling of the full half unit sphere of possible projection directions to be solvable. This necessitates the inclusion of a second tilt stage in the experimental set up and causes longer measurement times. Even with such a sample stage, a range of projection angles are usually obfuscated by the sample holder which leads to the so-called missing wedge problem in PF-TT due to which certain Fourier components of certain scattering directions cannot be probed\cite{schaff15, nielsen_2024_a}. While it can be shown that full angular sampling is necessary when different scattering directions are reconstructed independently\cite{schaff15}, it is less clear if this is necessary when using a model that enforces correlations between different scattering directions. This has lead several authors to assert that PF-TT is possible with such methods\cite{murer2021,zhao_2024} despite systematic studies generally showing this to have a negative impact on the quality of the reconstructions\cite{liebi18,carlsen_2024}.

As show here, ODF-TT is able to overcome the missing wedge problem of PF-TT and achieve solutions without a tilt rotation by utilizing the extra information given by the lattice symmetry and the sparse texture. 

\subsection{Analysis of reconstructions}\label{sec:analysis}

Each voxel of the reconstruction contain up to about 10 000 independent orientations and coefficients. While most coefficients are zero and this simplifies the analysis, both visualization and analysis of the reconstruction remain challenging. 

The main tool used in this paper is to select a main orientation for each voxel which is defined as the orientation corresponding to the basis function with the largest coefficient for each voxel. With this main orientation field we can compute inverse pole figure maps and orientations of crystallographic axes similar to what is done in EBSD analysis. The fraction of intensity attributable to this main orientation is computed by first selecting all orientations that fall within a 20$\si{\degree}$ radius of the main orientation using a symmetry equivalent distance metric. The fraction is then defined as the sum of all selected coefficients (including the main orientation itself) divided by the sum of all orientations.

Furthermore we use an approximate texture index defined as $J_{xyz} = \langle c_{xyz\mu}^2\rangle_\mu / \langle c_{xyz\mu}\rangle_\mu^2$. This effectively treats each mode as distinct orientation rather than a Gaussian distribution, but it has the same qualitative features as the exact texture index.

To quantify the twinning, we compute the misorientation between the main orientation and a secondary orientation, which is determined by the largest coefficient outside of a 20$\si{\degree}$ radius of the main orientation. Calling these two orientaions $g_1$ and $g_2$ respectively, the misorientation is defined as the orientation $g' = g_1g_s g_2^{-1}$ for $g_s$ in the symmetry group with minimum norm. The misorientation is the transformed using $g_m = g_s g' g_s^{-1}$ so the the axis of rotation falls within the fundamental region of the pole figure. 

One the misorientation is computed for all voxels with density above a threshold, histograms of the misorientation magnitude and misorientation axis are computed.

For the gastropod shell sample, the maps computed using the main orientation appear noisy. We therefore compute a mean orientation that also takes into account the weights of neighboring orientations. This is done by first selecting all orientation that fall within a 20$\si{\degree}$ radius of the main orientation. From these selected orientations a mean is computed by first rotating all orientations into a frame centered on the main orientation by a right handed application of the inverse of the main orientation, making sure to pick the smallest symmetry equivalent orientation for each. The transformed orientations are afterwards cast to Rodriguez vector space and a weighted arithmetic mean is computed of the vector components using the coefficients as weights. This mean orientation difference is afterwards transformed back into the original frame to give the mean orientation used for the plots.

\subsection{Experiment}

\subsubsection{Shot-peened martensite}

The first sample is a piece of martensitic steel with a cross section of about 0.8\,mm by 1\,mm. The sample was shot-peened from three sides in order to introduce high residual stresses and varying crystallite sizes from the edges to the center.

The experiment was carried out at the P21.2 beamline~\cite{Hegedus_2019} of PETRA III (Hamburg, Germany). A photon energy of 68\,keV was selected by a double Laue monochromator. Compound refractive lenses shaped a pencil beam of $8~\si{\micro m}$ width by $3~\si{\micro m}$ height with a photon flux of about $10^{10}$ photons per second. A VAREX XRD4343CT detector with a pixel size of $150\si{\micro m}$ was positioned about 0.8\,m downstream of the sample and provided diffraction angles of up to $2\theta_{max}=14\si{\degree}$. 

The single rotation axis was oriented in vertical direction. The sample was scanned with stepwise translations orthogonal to the rotation axis and continuous rotation. In total, 200 projections were collected over $360\si{\degree}$ and the sample was translated 200 times in 8\,$\si{\micro m}$ steps, which resulted in a field of view of 1.6\,mm. The exposure time was 0.4\,s and a total number of 40 000 frames were acquired, which took approximately 5\,h.

The integrated intensity was determined for each of the seven diffraction rings that were fully captured by the detector and azimuthally regrouped into 72 equally spaced azimuthal bins. The resulting sinograms were corrected by the absorption signal provided by a beamstop diode.

The reconstruction was performed without regularization by running the algorithm for 1000 iterations which took 8 hours to complete. The 2D slice was reconstructed on a 150$\times$150 pixel grid with an ODF expansion using 10 000 orientations in the asymmetric zone. For each orientation a spherically symmetric Gaussian with $\sigma = 0.05\mathrm{rad}$ was used. This gives 15 million data points and 240 million degrees of freedom of which 0.6 million were non-zero in the converged solution. The final reconstruction has an $R^2$ value of 0.8. 

\subsubsection{Gastropod shell}

The second sample consists of a piece of the shell of a \textit{helix pomatia} (Roman snail) which was found, already dead and without traces of the soft body remaining, in the forests near Beznau, Aargau, Switzerland. A small piece measuring 3\,mm by 3\,mm by 2\,mm of the columella was taken about 5\,mm below the apex (See Fig. \ref{fig:snail_1}b)). The snail shell consists mainly of aragonite and no other crystalline components could be identified in the diffraction patterns.

The sample was measured at the cSAXS beamline at the Swiss Light Source. The experiments were performed with a photon energy of $18\,\si{keV}$ and a sample-to-detector distance of $0.2\,\si{m}$ using a Pilatus 2M detector. The beam was focused to a spot of approximately $50\,\si{\micro m}\times50\,\si{\micro m}$ and the sample was raster-scanned in 2D through the focused beam with continuous translations in one direction and a step-size of $50\,\si{\micro m}$ and step-scanning with two orthogonal rotation stages. The intensity of the direct bream was measured on the image-detector behind a semi-transparent beamstop constructed of several small discs of single-crystal silicon glued together.

For this sample, a full TT-dataset was measured, containing measurement made at tilt angles extending up to 45\si{\degree}. This allows the texture to be reconstructed with the more established PF-TT approach and the two approaches to be compared as well as a comparison between using the full data set and using a zero-tilt geometry only.

The diffraction patterns contain 16 rings that are fully covered by the detector but due to partial overlaps of some peaks, only eight rings were used for the reconstructions. These rings were azimuthally regrouped into 96 equally spaced bins.

For the reconstructions, we use a grid of 8 000 orientations and Gaussian functions with width $\sigma = 0.1\,\mathrm{rad}$ for the basis set. The gastropod shell consists of aragonite of the orthohombic crystal class which means the rotation group contains only 4 symmetries compared to the 24 in the cubic crystal class of ferrite. Therefore, even though the number of grid orientations are similar, the angular resolution of the gastropod shell reconstruction is about a factor of two worse than the martensite reconstruction. The full three dimensional problem has 600 million data points hereof 150 million in the zero tilt part and the model uses 1.8 billion degrees of freedom.

\bibliography{bib}

\begin{thebibliography}{10}

\bibitem{JUULJENSEN2020}
D.~{Juul Jensen} and Y.B. Zhang.
\newblock Impact of 3d/4d methods on the understanding of recrystallization.
\newblock {\em Current Opinion in Solid State and Materials Science},
  24(2):100821, 2020.

\bibitem{Knipschildt_2023}
E~F~F Knipschildt, Y~B Zhang, T~Yu, X~Lei, R~E Sanders, and D~Juul Jensen.
\newblock Artifacts of particle stimulated nucleation in 2d and 3d—a
  numerical analysis.
\newblock {\em Journal of Physics: Conference Series}, 2635(1):012030, nov
  2023.

\bibitem{harding_xrd_1987}
G.~Harding, J.~Kosanetzky, and U.~Neitzel.
\newblock X-ray diffraction computed tomography.
\newblock {\em Medical Physics}, 14(4):515--525, 1987.

\bibitem{stock08}
SR. Stock, F.~{De Carlo}, and JD. Almer.
\newblock High energy x-ray scattering tomography applied to bone.
\newblock {\em Journal of Structural Biology}, 161(2):144--150, 2008.

\bibitem{bleuet08}
J.~Bleuet, E.~Welcomme, E.~Dooryhée, J.~Susini, JL. Hodeau, and P.~Walter.
\newblock Probing the structure of heterogeneous diluted materials by
  diffraction tomography.
\newblock {\em Nature Materials}, 7:468--472, 2008.

\bibitem{malecki14}
A.~Malecki, G.~Potdevin, T.~Biernath, E.~Eggl, K.~Willer, T.~Lasser,
  J.~Maisenbacher, J.~Gibmeier, A.~Wanner, and F.~Pfeiffer.
\newblock X-ray tensor tomography(a).
\newblock {\em Europhysics Letters}, 105:38002, 2014.

\bibitem{liebi15}
M.~Liebi, M.~Georgiadis, A.~Menzel, P~Schneider, J.~Kohlbrecher, O~Bunk, and
  M.~Guizar-Sicairos.
\newblock Nanostructure surveys of macroscopic specimens by small-angle
  scattering tensor tomography.
\newblock {\em Nature}, 527:349--352, 2015.

\bibitem{schaff15}
F.~Schaff, M.~Bech, P.~Zalansky, C.~Jud, M.~Liebi, M.~Guizar-Sicairon, and
  F.~Pfeiffer.
\newblock Six-dimensional real and reciprocal space small-angle x-ray
  scattering tomography.
\newblock {\em Nature}, 527:353--356, 2015.

\bibitem{frewein_2024}
M.~P.~K. Frewein, J.~Mason, B.~Maier, H.~C{\"{o}}lfen, A.~Medjahed,
  M.~Burghammer, M.~Allain, and T.~A. Gr{\"{u}}newald.
\newblock {Texture tomography, a versatile framework to study crystalline
  texture in 3D}.
\newblock {\em IUCrJ}, 11(5), Sep 2024.

\bibitem{hayashi15}
Y.~Hayashi, Y.~Hirose, and Y.~Seno.
\newblock Polycrystal orientation mapping using scanning three-dimensional
  x-ray diffraction microscopy.
\newblock {\em Journal of Applied Crystallography}, 48:1094--1101, 2015.

\bibitem{henningsson2020}
N.~A. Henningsson, S.~A. Hall, J.~P. Wright, and J.~Hektor.
\newblock Reconstructing intragranular strain fields in polycrystalline
  materials from scanning 3dxrd data.
\newblock {\em Journal of Applied Crystallography}, 53(2):314--325, Apr 2020.

\bibitem{bucsek_2023}
W.~Li, H.~Sharma, P.~Kenesei, S.~Ravi, H.~Sehitoglu, and A.~Buscek.
\newblock Resolving intragranular stress fields in plastically deformed
  titanium using point-focused high-energy diffraction microscopy.
\newblock {\em Journal of Materials Research}, 2023.

\bibitem{poulsen04}
H.~Poulsen.
\newblock {\em Three-Dimensional X-ray Diffraction Microscopy}.
\newblock Springer Berlin, Heidelberg, 2004.

\bibitem{Poulsen_2001}
H.~F. Poulsen, S.~F. Nielsen, E.~M. Lauridsen, S.~Schmidt, R.~M. Suter,
  U.~Lienert, L.~Margulies, T.~Lorentzen, and D.~Juul~Jensen.
\newblock {Three-dimensional maps of grain boundaries and the stress state of
  individual grains in polycrystals and powders}.
\newblock {\em Journal of Applied Crystallography}, 34(6):751--756, Dec 2001.

\bibitem{suter_2006}
R.~M. Suter, D.~Hennessy, C.~Xiao, and U.~Lienert.
\newblock {Forward modeling method for microstructure reconstruction using
  x-ray diffraction microscopy: Single-crystal verification}.
\newblock {\em Review of Scientific Instruments}, 77(12):123905, 12 2006.

\bibitem{Schmidt_2008}
S.~Schmidt, U.L. Olsen, H.F. Poulsen, H.O. Sørensen, E.M. Lauridsen,
  L.~Margulies, C.~Maurice, and D.~{Juul Jensen}.
\newblock Direct observation of 3-d grain growth in al–0.1
\newblock {\em Scripta Materialia}, 59(5):491--494, 2008.

\bibitem{Ludwig_2008}
Wolfgang Ludwig, S{\o}eren Schmidt, Erik~Mejdal Lauridsen, and Henning~Friis
  Poulsen.
\newblock {X-ray diffraction contrast tomography: a novel technique for
  three-dimensional grain mapping of polycrystals. I. Direct beam case}.
\newblock {\em Journal of Applied Crystallography}, 41(2):302--309, Apr 2008.

\bibitem{vigano_2016}
N.~Vigano, A.~Tanguy, S.~Hallais, A.~Dimanov, M.~Bornert, K.~J. Batenburg, and
  W.~Ludwig.
\newblock Three-dimensional full-field x-ray orientation microscopy.
\newblock {\em Scientific Reports}, 2016.

\bibitem{oddershede_2019}
J.~Oddershede, J.~Sun, N.~Gueninchault, F.~Bachmann, H.~Bale, C.~Holzner, and
  E.~Lauridsen.
\newblock Non-destructive characterization of polycrystalline materials in 3d
  by laboratory diffraction contrast tomography.
\newblock {\em Integrating Materials and Manufacturing Innovation}, 2019.

\bibitem{gruenewald20}
TA. Gr{\"{u}}newald, M.~Liebi, NK. Wittig, A.~Johannes, T.~Sikjaer,
  L.~Rejnmark, Z.~Gao, M.~Rosenthal, M.~Guizar-Sicairos, H.~Birkedal, and
  M.~Burghammer.
\newblock Mapping the 3d orientation of nanocrystals and nanostructures in
  human bone: Indications of novel structural features.
\newblock {\em Science Advances}, 6:eaba4171, 2020.

\bibitem{gruenewald23}
TA. Gr{\"{u}}newald, A.~Johannes, NK. Wittig, J.~Palle, A.~Rack, M.~Burghammer,
  and H.~Birkedal.
\newblock Bone mineral properties and 3d orientation of human lamellar bone
  around cement lines and the haversian system.
\newblock {\em IUCrJ}, 10:189--198, 2023.

\bibitem{carlsen_2024}
Mads Carlsen, Christian Appel, William Hearn, Martina Olsson, Andreas Menzel,
  and Marianne Liebi.
\newblock {X-ray tensor tomography for small-grained polycrystals with strong
  texture}.
\newblock {\em Journal of Applied Crystallography}, 57(4):986--1000, Aug 2024.

\bibitem{nielsen_2024_a}
L.~C. Nielsen, T~Tänzer, I~Rodriguez-Fernandez, P~Erhart, and M~Liebi.
\newblock Investigating the missing wedge problem in small-angle x-ray
  scattering tensor tomography across real and reciprocal space.
\newblock {\em Journal of Synchrotron Radiation}, 31(5), 2024.

\bibitem{Kurdjumow_1930Z}
G.~{Kurdjumow} and G.~{Sachs}.
\newblock {{\"U}ber den Mechanismus der Stahlh{\"a}rtung}.
\newblock {\em Zeitschrift fur Physik}, 64(5-6):325--343, May 1930.

\bibitem{guo_2004}
Z.~Guo, C.S. Lee, and J.W. Morris.
\newblock On coherent transformations in steel.
\newblock {\em Acta Materialia}, 52(19):5511--5518, 2004.

\bibitem{boggild_1930}
Bøggild~O. B.
\newblock The shell structure of the mollusks.
\newblock {\em Det Kongelige Danske Videnskabernes Selskabs Skrifter.
  Naturvidenskabelig og Mathematisk Afdeling, Raekke 9}, 2:231--326, 1930.

\bibitem{nielsen2023b}
L.C. Nielsen, M.~Carlsen, M.~Liebi, and P.~Erhart.
\newblock mumott - a python library for the analysis of photon probe tensor
  tomography data., 2023.

\bibitem{hayashi_2023}
Yujiro Hayashi, Daigo Setoyama, Kunio Fukuda, Katsuharu Okuda, Naoki Katayama,
  and Hidehiko Kimura.
\newblock Scanning three-dimensional x-ray diffraction microscopy with a spiral
  slit.
\newblock {\em Quantum Beam Science}, 7(2), 2023.

\bibitem{henningsson_2024}
Axel Henningsson, Mustafacan Kutsal, Jonathan~P. Wright, Wolfgang Ludwig,
  Henning~Osholm Sørensen, Stephen~A. Hall, Grethe Winther, and Henning~F.
  Poulsen.
\newblock Microstructure and stress mapping in 3d at industrially relevant
  degrees of plastic deformation, 2024.

\bibitem{hayashi_2023_b}
Yujiro Hayashi and Hidehiko Kimura.
\newblock Scanning three-dimensional x-ray diffraction microscopy for carbon
  steels.
\newblock {\em Quantum Beam Science}, 7(3), 2023.

\bibitem{Vigano2024}
Nicola Vigan{\`{o}}, Wolfgang Ludwig, and Kees~Joost Batenburg.
\newblock {Reconstruction of local orientation in grains using a discrete
  representation of orientation space}.
\newblock {\em Journal of Applied Crystallography}, 47(6):1826--1840, Dec 2014.

\bibitem{simons_2015}
H.~Simons, A.~King, W.~Ludwig, C.~Detlefs, W.~Panteleon, S.~Schmidt,
  F.~St\"ohr, I.~Snigreva, A.~Snigrev, and H.~F. Poulsen.
\newblock Dark-field x-ray microscopy for multiscale structural
  characterization.
\newblock {\em Nature Communications}, 2015.

\bibitem{yang2004}
W.~Yang, B.C. Larson, J.Z. Tischler, G.E. Ice, J.D. Budai, and W.~Liu.
\newblock Differential-aperture x-ray structural microscopy: a
  submicron-resolution three-dimensional probe of local microstructure and
  strain.
\newblock {\em Micron}, 35(6):431--439, 2004.
\newblock International Wuhan Symposium on Advanced Electron Microscopy.

\bibitem{Larson_2013}
B.~C. Larson and L.~E. Levine.
\newblock {Submicrometre-resolution polychromatic three-dimensional X-ray
  microscopy}.
\newblock {\em Journal of Applied Crystallography}, 46(1):153--164, Feb 2013.

\bibitem{Yildirim_2023}
Can Yildirim, Carsten Detlefs, Albert Zelenika, Henning~F. Poulsen, Raquel
  Rodriguez-Lamas, Philip~K. Cook, Mustafacan Kutsal, and Nikolas Mavrikakis.
\newblock Exploring 4d microstructural evolution in a heavily deformed ferritic
  alloy.
\newblock {\em Journal of Physics: Conference Series}, 2635(1):012040, nov
  2023.

\bibitem{yu_2023}
Tianbo Yu, Chuanshi Hong, Yubin Zhang, Adam Lindkvist, Wenjun Liu, Jon
  Tischler, and Dorte~Juul Jensen.
\newblock Recovery of deformation microstructure in the bulk interior revealed
  by synchrotron x-ray micro-diffraction.
\newblock {\em Materials Characterization}, 202:112997, 2023.

\bibitem{Niessen_2022}
Frank Niessen, Tuomo Nyyss{\"{o}}nen, Azdiar~A. Gazder, and Ralf Hielscher.
\newblock {Parent grain reconstruction from partially or fully transformed
  microstructures in {\it MTEX}}.
\newblock {\em Journal of Applied Crystallography}, 55(1):180--194, Feb 2022.

\bibitem{nielsen2023}
L.C. Nielsen, P.~Erhart, M.~Guizar-Sicairos, and M.~Liebi.
\newblock Small-angle scattering tensor tomography algorithm for robust
  reconstruction of complex textures.
\newblock {\em Acta Crystallographica Section A}, 79(6):515--526, 2023.

\bibitem{roe_1965}
Ryong‐Joon Roe.
\newblock {Description of Crystallite Orientation in Polycrystalline Materials.
  III. General Solution to Pole Figure Inversion}.
\newblock {\em Journal of Applied Physics}, 36(6):2024--2031, 06 1965.

\bibitem{Bunge_book_1}
H.-J. (Hans~Joachim) Bunge.
\newblock {\em Mathematische Methoden der Texturanalyse : Mit 86 Abbildungen
  und 31 Tabellen, sowie 7 Abbildungen und 9 Tabellen im Anhang}.
\newblock Akademie-Verlag, 1969.

\bibitem{matthies_1979}
S.~Matthies.
\newblock On the reproducibility of the orientation distribution function of
  texture samples from pole figures (ghost phenomena).
\newblock {\em physica status solidi (b)}, 92(2):K135--K138, 1979.

\bibitem{ruer_1977}
D.~Ruer and R.~Baro.
\newblock Méthode vectorielle d'analyse de la texture des matériaux
  polycristallins de réseau cubique.
\newblock {\em Journal of Applied Crystallography}, 10(6):458--464, 1977.

\bibitem{schaeben_1994}
H.~Schaeben.
\newblock Analogy and duality of texture analysis by harmonics or indicators.
\newblock {\em Journal of Scientific Computing}, 1994.

\bibitem{barton_2002}
Donald E.~Boyce Nathan R.~Barton and Paul~R. Dawson.
\newblock Pole figure inversion using finite elements over rodrigues space.
\newblock {\em Textures and Microstructures}, 35(2):113--144, 2002.

\bibitem{schaeben_1996}
H.~Schaeben.
\newblock A unified view of methods to resolve the inverse problem of texture
  goniometry.
\newblock {\em Textures and Microstructures}, 1996.

\bibitem{schaeben_1988}
H.~Schaeben.
\newblock Entropy optimization in texture goniometry. i. methodology.
\newblock {\em physica status solidi (b)}, 148(1):63--72, 1988.

\bibitem{matthies_1982}
S.~Matthies.
\newblock {\em Aktuelle Probleme der quantitativen Texturanalyse}.
\newblock Akademie der Wissenschaften der DDR, Dresden, Dresden, 1972.

\bibitem{dahms_1988}
M.~Dahms and H.~J. Bunge.
\newblock A positivity method for the determination of complete orientation
  distribution functions.
\newblock {\em Textures and Microstructures}, 1988.

\bibitem{hielscher_2008}
R.~Hielscher and H.~Schaeben.
\newblock {A novel pole figure inversion method: specification of the {\it
  MTEX} algorithm}.
\newblock {\em Journal of Applied Crystallography}, 41(6):1024--1037, Dec 2008.

\bibitem{murer2021}
F.~K. M{\"{u}}rer, A.~S. Madathiparambil, K.~R. Tekseth, M.~Di Michiel,
  P.~Cerasi, B.~Chattopadhyay, and D.~W. Breiby.
\newblock Orientational mapping of minerals in pierre shale using x-ray
  diffraction tensor tomography.
\newblock {\em IUCrJ}, 8(5):747--756, 2021.

\bibitem{zhao_2024}
Xiaoyi Zhao, Zheng Dong, Chenglong Zhang, Himadri Gupta, Zhonghua Wu, Wenqiang
  Hua, Junrong Zhang, Pengyu Huang, Yuhui Dong, and Yi~Zhang.
\newblock {A step towards 6D WAXD tensor tomography}.
\newblock {\em IUCrJ}, 11(4), Jul 2024.

\bibitem{liebi18}
M.~Liebi, M.~Georgiadis, J.~Kohlbrecher, M.~Holler, J.~Raabe, I.~Usov,
  A.~Menzel, P.~Schneider, O.~Bunk, and M.~Guizar-Sicairos.
\newblock Small-angle x-ray scattering tensor tomography: model of the
  three-dimensional reciprocal-space map, reconstruction algorithm and angular
  sampling requirements.
\newblock {\em Acta Crystallographica Section A}, 74:12--24, 2018.

\bibitem{Hegedus_2019}
Z~Heged\"us, T~Müller, J~Hektor, E~Larsson, T~Bäcker, S~Haas, ALC Conceiçao,
  S~Gutschmidt, and U~Lienert.
\newblock Imaging modalities at the swedish materials science beamline at petra
  iii.
\newblock {\em IOP Conference Series: Materials Science and Engineering},
  580(1):012032, aug 2019.

\end{thebibliography}
\bibliographystyle{unsrt}

\newpage

\renewcommand{\figurename}{Supplementary Figure} 
\renewcommand{\thesection}{A\arabic{section}}
\renewcommand{\thetable}{A\arabic{table}}
\renewcommand{\thefigure}{A\arabic{figure}}
\setcounter{figure}{0} 

\onecolumn
\section{Supplementary}
\FloatBarrier

\subsection{Importance of the non-negativity constraint}
\begin{figure}[H]
    \centering
    \includegraphics[width = 0.85\columnwidth]{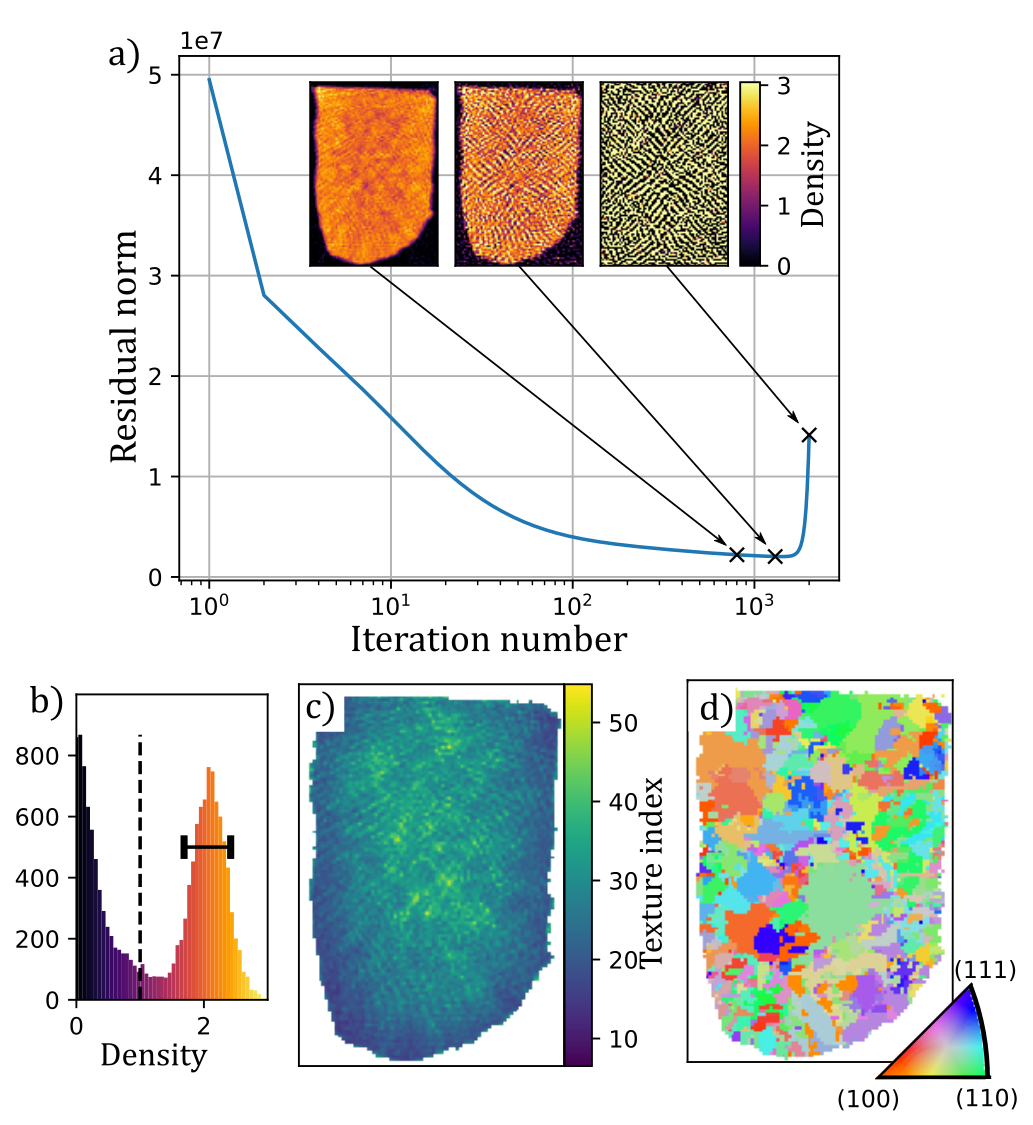}
    \caption{Tomograms of the martensite reconstruction without utilizing the non-negativity constraint. a) shows a convergence cuve and the reconstructed density at three different point during the optimization and b) displays a histogram of the density values of the 800th iterate. b) shows the corresponding calculated texture index and d) shows a inverse pole figure map of the main orientation.}
    \label{fig:supplement_1}
\end{figure}

To illustrate the importance of the non-negativity constraint, we performed a reconstruction of the martensite data without using the non-negativity constraint. Figure \ref{fig:supplement_1} shows the results of this reconstruction. It is apparent, that the reconstruction without the non-negativity constraints suffers from a high-frequency reconstruction artifact that grows exponentially in amplitude after about 1000 iterations. The fixed step-size algorithm used here is not stable for this problem, but a reasonable reconstruction can be achieved by stopping the optimization procedure early.

\FloatBarrier
\newpage
\FloatBarrier
\subsection{Comparison of pole-figures}
\begin{figure}[H]
    \centering
    \includegraphics[width = 0.5\columnwidth]{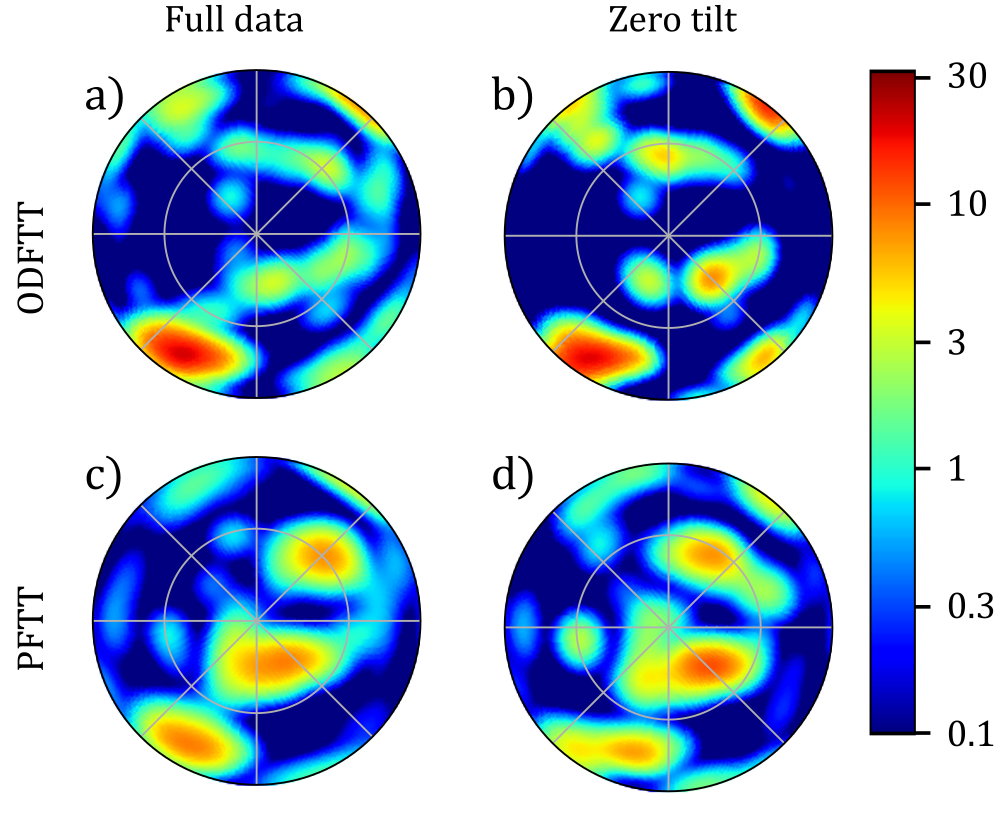}
    \caption{Pole-figures of a single voxel of the snail-sample with four different reconstructions.}
    \label{fig:supplement_3}
\end{figure}

To give more insight into the comparison of ODF-TT and PF-TT, we plot the \{200\} pole figure of a single voxel of the reconstructions of the gastropod shell data in Fig. \ref{fig:supplement_3}. It appears that the PF-TT resonstruction do reconstruct significant scattering in the direction where ODF-TT has the majority of the intensity. However, the PF-TT reconstruction also has significant intensity in two other directions with similar amplitude.

\subsection{L1 regularization}

\begin{figure}[H]
    \centering
    \includegraphics[width = 0.5\columnwidth]{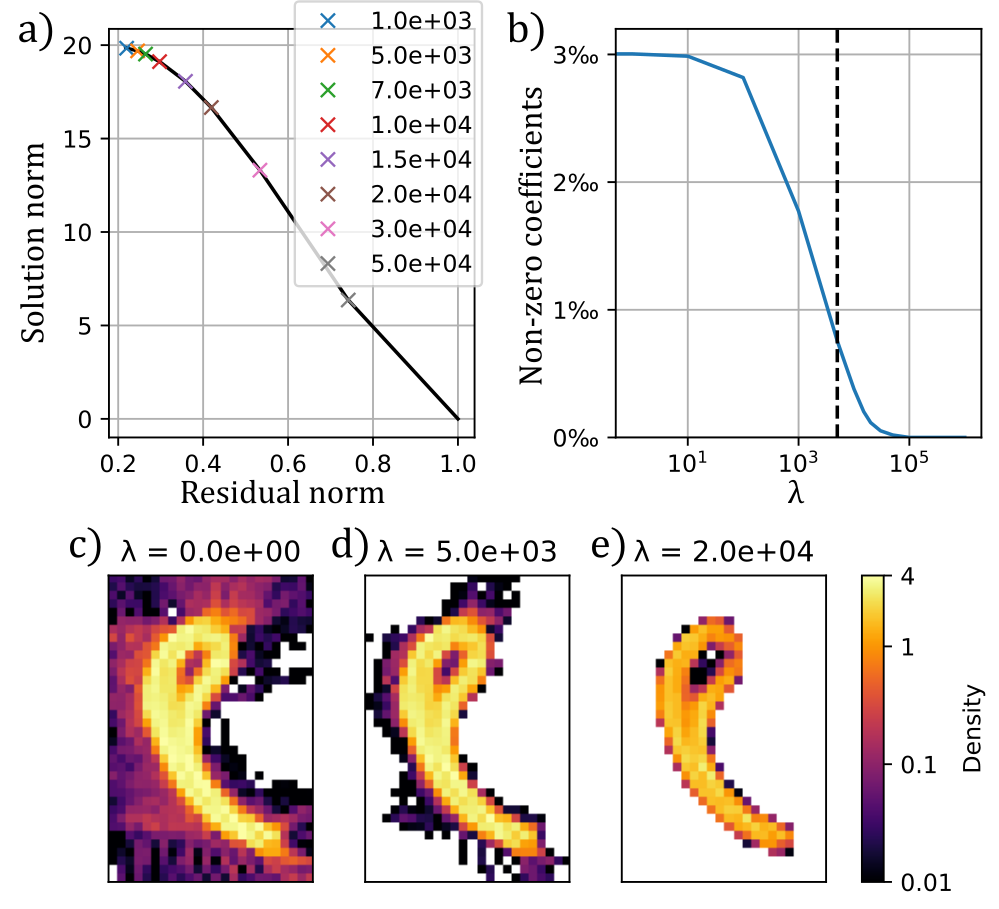}
    \caption{Effect of regularization. a) Shows how the residual-norm ($|\mathbf{I}-\mathrm{A}\mathbf{x}|^2$) and solution-norm ($|\mathbf{x}|^2$) after 500 iterations depends on the choice of the regularization parameter $\lambda$. b) Percentage of non-zero coefficients as a function of regularization parameter. At $\lambda = 5\cdot 10^{3}$ (dashed line) the number of non-zero coefficients have been reduced by about a factor of 4. Plots of the reconstructed density at c) $\lambda = 0.0$, d) $\lambda = 5\cdot 10^{3}$, and e) $\lambda = 2\cdot 10^{4}$. White regions in these images correspond to voxels where all coefficients are zero.}
    \label{fig:supplement_4}
\end{figure}

For the samples investigated in this paper, good reconstructions can be achieved without use of L1 regularization. There are however advantages to using the regularization. As seen in Fig. \ref{fig:supplement_4}, using a regularized reconstruction can reduce streaking artifacts in the reconstruction and significantly decrease the number of non-zero coefficients. The latter is useful as it makes the reconstructed volumes easier to store and process for further analysis.

\subsection{Residuals}

\begin{figure}[H]
    \centering
    \includegraphics[width = 0.75\columnwidth]{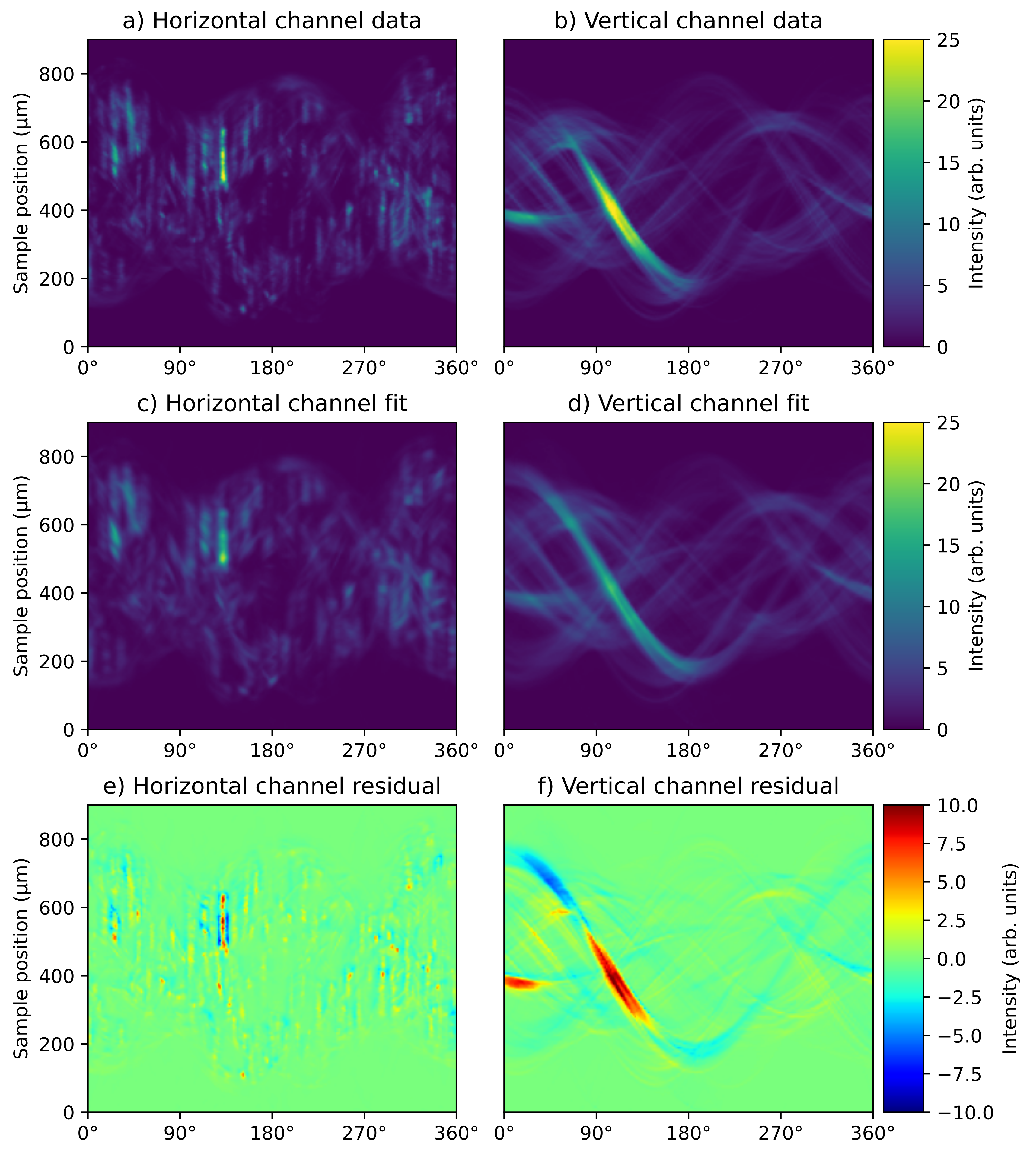}
    \caption{Comparison of data and fit of the martensite reconstruction. a,b) experimental sinograms of the 110 peak for a single azimuthal detector channel. c,d) fitted sinograms of the same angular channel. e,f) residual. a,c,e) show an angular channel orthogonal to the rotation axis and b,d,f) show an angular channel parallel to the rotation axis. }
    \label{fig:suppl_residual}
\end{figure}

To get some insight into the quality of the fit and potential source of errors, we show the fit and residuals in Fig. \ref{fig:suppl_residual}. The magnitude of the residual is lower than 2 throughout most of both sinograms, but peaks at a large value of around $\pm 10$ near the most intense features in each sinogram. We believe that these large errors are due to the limited angular resolution of the reconstruction. The two sinograms displayed here were picked arbitrarily. The full dataset contains 504 such sinograms.

\end{document}